\documentclass[nofootinbib,prd]{revtex4}%
\usepackage{amsmath}
\usepackage{amsfonts}
\usepackage{amssymb}
\usepackage{graphicx}%
\begin{document}
\title{Particle propagation and effective space-time in Gravity's Rainbow}
\author{Remo Garattini}
\email{Remo.Garattini@unibg.it}
\affiliation{Universit\`{a} degli Studi di Bergamo, Facolt\`{a} di Ingegneria,}
\affiliation{Viale Marconi 5, 24044 Dalmine (Bergamo) Italy}
\affiliation{I.N.F.N. - sezione di Milano, Milan, Italy.}
\author{Gianluca Mandanici}
\email{Gianluca.Mandanici@unibg.it}
\affiliation{Universit\`{a} degli Studi di Bergamo, Facolt\`{a} di Ingegneria, Viale
Marconi 5, 24044 Dalmine (Bergamo) Italy.}

\begin{abstract}
Basing on the results obtained in a our previous study on Gravity's Rainbow,
we determine the quantum corrections to the space-time metric for the
Schwarzschild and the de Sitter background, respectively. We analyze how
quantum fluctuations alter these metrics inducing modifications on the
propagation of test particles. Significantly enough we find that quantum
corrections can become relevant not only for particles approaching the Planck
energy but, due to the one loop contribution, even for low-energy particles as
far as Planckian length scales are considered. We briefly compare our results
with others obtained in similar studies and with the recent experimental OPERA
announcement of superluminal neutrino propagation.

\end{abstract}
\maketitle

\section{Introduction}

It is almost more than a decade that the idea of modifying\textbf{ }the
Lorentz-Poincar\'{e} symmetry at the Planck scale has being considered in a
systematic way. The pioneering idea is known under the name of \textit{Doubly
Special Relativity}, according to which the modifications of the
Lorentz-Poincar\'{e} symmetry should occur also preserving the relativity
principle, i.e. preserving the equivalence of all inertial observers
\cite{AmelinoCamelia:2000mn,AmelinoCamelia:2000ge,Magueijo:2001cr,KowalskiGlikman:2004qa}%
. Actually, in the most studied doubly-special-relativity proposals, the
Lorentz sector of the Poincar\'{e} symmetry is modified only for its action on
the energy-momentum space of the test particles. This action in fact becomes
nonlinear and Planck-scale dependent. The resulting deformed symmetry produces
deformed Casimirs and, as a consequence, deformed energy-momentum dispersion
relations of the type%
\begin{equation}
E^{2}g_{1}^{2}\left(  E/E_{P}\right)  -p^{2}g_{2}^{2}\left(  E/E_{P}\right)
=m^{2},\label{mdisp}%
\end{equation}
where $g_{1}\left(  E/E_{P}\right)  $ and $g_{2}\left(  E/E_{P}\right)  $ are
two arbitrary functions whose contributions become relevant only at Planckian
energies. In order to reproduce the usual behavior at low energies we must
have%
\begin{equation}
\lim_{E/E_{P}\rightarrow0}g_{1}\left(  E/E_{P}\right)  =1\qquad\text{and}%
\qquad\lim_{E/E_{P}\rightarrow0}g_{2}\left(  E/E_{P}\right)  =1.\label{lim}%
\end{equation}
In \cite{Magueijo:2002xx} these type of relativistic symmetries have been
applied to general relativity. The resulting gravity model, known as Gravity's
Rainbow, produces a correction to the space-time metric that becomes
significative as soon as the particle energy/momentum approaches the Planck
energy/momentum, being otherwise these quantum corrections Planck-scale
suppressed. In a recent paper \cite{GaMa1} we have discussed the modifications
induced by the Gravity's Rainbow\textbf{ }of \cite{Magueijo:2002xx} on the
zero-point gravitational energy. In particular it has been shown that certain
classes of deformed dispersion relations lead to a finite Zero Point Energy
(ZPE) for the gravitational field and induce a finite cosmological constant
(see also
\cite{DeWitt:1967yk,Garattini:2004zu,Garattini:2005kx,Garattini:2005ky}),
avoiding therefore the traditional procedures of regularization and
renormalization. Procedures that can be avoided even in Noncommutative
geometry\cite{RGPN}. An interesting aspect is that from the Einstein's Field
Equations, $G_{\mu\nu}=8\pi G\,T_{\mu\nu}$ (with $c=1$), written in an
orthonormal reference frame for a spherically symmetric space-time%
\begin{equation}
ds^{2}=-\exp[-2\Phi(r)]\frac{dt^{2}}{g_{1}^{2}\left(  E/E_{P}\right)  }%
+\frac{dr^{2}}{\left(  1-\frac{b\left(  r\right)  }{r}\right)  g_{2}%
^{2}\left(  E/E_{P}\right)  }+\frac{r^{2}}{g_{2}^{2}\left(  E/E_{P}\right)
}\left(  d\theta^{2}+\sin^{2}\theta d\phi^{2}\right)  ,\label{metric}%
\end{equation}
we obtain the following set of equations%
\begin{align}
\rho(r) &  =\frac{1}{8\pi G}\;\frac{b^{\prime}}{r^{2}}\,,\label{rhoWH}\\
p_{r}(r) &  =\frac{1}{8\pi G}\left[  2\left(  1-\frac{b}{r}\right)  \frac
{\Phi^{\prime}}{r}-\frac{b}{r^{3}}\right]  \,,\label{prWH}\\
p_{t}(r) &  =\frac{1}{8\pi G}\left(  1-\frac{b}{r}\right)  \left[
\Phi^{\prime\prime}+(\Phi^{\prime})^{2}\right.  \nonumber\\
&  \left.  -\frac{b^{\prime}r-b}{2r(r-b)}\Phi^{\prime}-\frac{b^{\prime}%
r-b}{2r^{2}(r-b)}+\frac{\Phi^{\prime}}{r}\right]  \,.\label{ptWH}%
\end{align}
$\Phi(r)$ is termed the redshift function, while $b\left(  r\right)  $ is the
shape function subject to the only condition $b\left(  r_{t}\right)  =r_{t}$.
$\rho(r)$ is the energy density, $p_{r}(r)$ is the radial pressure, and
$p_{t}(r)$ is the lateral pressure measured in the orthogonal direction to the
radial direction. Among these equations if we focus our attention on
Eq.$\left(  \ref{rhoWH}\right)  $ and we impose that%
\begin{equation}
\rho(r)=\frac{\Lambda(r)}{8\pi G},\label{rho}%
\end{equation}
where $\Lambda(r)$ is the cosmological constant defined by Eq.$\left(
\ref{EWDW}\right)  $, we find that it must be%
\begin{equation}
b_{qc}(r)=b_{qc}(+\infty)+\int_{+\infty}^{r}\Lambda(r^{\prime})r^{\prime
2}dr^{\prime}.\label{diffbr}%
\end{equation}
Therefore, we expect that the quantum-induced cosmological constant obtained
in Ref.\cite{GaMa1}, can be considered as an energy density source leading to
an effective metric%
\begin{equation}
b_{eff}(r)=b_{cl}(r)+b_{qc}(r),\label{effbr}%
\end{equation}
where $b_{cl}(r)$ is the classical shape function, and where $b_{qc}(r)$
accounts for the quantum corrections. In this paper we extend the analysis of
Ref.\cite{GaMa1} to find the distortion of the classical metric caused by
quantum fluctuations. Such a distortion will produce effects on the motion of
some test particle in the considered background which, in principle can be
measured. The paper is organized as follows. In section \ref{S2} we outline
the procedure we used to calculate the effective metric from the quantum
induced cosmological fluctuations. In section \ref{S3} we study the (quantum)
induced space-time metric for a Schwarzschild background. In section \ref{S4}
we discuss the implications of the Schwarzschild effective metric on a
orbiting test particle. In section \ref{S5} we analyze the effects of the
quantum corrections on a de Sitter space-time, briefly discussing possible
implications for particles propagating on a de Sitter metric. In Section
\ref{S6} we discuss our conclusions.

\section{Induced cosmological constant and effective metric}

\label{S2}The procedure followed in this work relies heavily on the formalism
outlined in Ref.\cite{GaMa1}, where the graviton one loop contribution to the
cosmological constant in a background of the form $\mathbf{(}\ref{metric}%
\mathbf{)}$ was computed. Rather than reproduce the formalism, we shall refer
the reader to Ref.\cite{GaMa1} for details, when necessary. However, for
self-completeness and self-consistency, we present here a brief outline of the
formalism used. In this paper, rather than working with energy density as in
Ref.\cite{GaMa1}, we shall integrate over the whole hypersurface $\Sigma$ to
obtain enough information for the effective metric. The idea is to consider
the Wheeler-De Witt equation as a formal eigenvalue equation where the
cosmological constant divided by the Newton's constant is the desired
eigenvalue. The expectation value we are interested is $\left(  \kappa=8\pi
G\right)  $%
\begin{equation}
\frac{\left\langle \Psi\left\vert \int_{\Sigma}d^{3}x\Lambda_{\Sigma
}\right\vert \Psi\right\rangle }{V\left\langle \Psi|\Psi\right\rangle }%
=-\frac{\Lambda_{c}}{\kappa},\label{EWDW}%
\end{equation}
where%
\begin{equation}
\Lambda_{\Sigma}=\left(  2\kappa\right)  G_{ijkl}\pi^{ij}\pi^{kl}%
\mathcal{-}\frac{\sqrt{g}R}{\left(  2\kappa\right)  g_{2}\left(  E\right)
}\!{}\!\label{LWDW}%
\end{equation}
and $V$ is the volume. However if we use the line element $\left(
\ref{metric}\right)  $, the expectation value transforms into%
\begin{equation}
\frac{g_{2}^{3}\left(  E\right)  }{\tilde{V}}\frac{\left\langle \Psi\left\vert
\int_{\Sigma}d^{3}x\tilde{\Lambda}_{\Sigma}\right\vert \Psi\right\rangle
}{\left\langle \Psi|\Psi\right\rangle }=-\frac{\Lambda_{c}}{\kappa
},\label{WDW3}%
\end{equation}
with%
\begin{equation}
\tilde{\Lambda}_{\Sigma}=\left(  2\kappa\right)  \frac{g_{1}^{2}\left(
E\right)  }{g_{2}^{3}\left(  E\right)  }\tilde{G}_{ijkl}\tilde{\pi}^{ij}%
\tilde{\pi}^{kl}\mathcal{-}\frac{\sqrt{\tilde{g}}\tilde{R}}{\left(
2\kappa\right)  g_{2}\left(  E\right)  }\!{}\!,\label{LambdaR}%
\end{equation}
where%
\begin{equation}
G_{ijkl}=\frac{1}{2\sqrt{g}}\left(  g_{ik}g_{jl}+g_{il}g_{jk}-g_{ij}%
g_{kl}\right)  =\frac{\tilde{G}_{ijkl}}{g_{2}\left(  E\right)  }.
\end{equation}
With the help of Gaussian trial wave functionals, a solution of Eq.$\left(
\ref{WDW3}\right)  $ has been obtained in Ref.\cite{GaMa1} to one loop, whose
form is%
\begin{equation}
\frac{\Lambda}{8\pi G}=-\frac{1}{3\pi^{2}}\sum_{i=1}^{2}\int_{E^{\ast}%
}^{+\infty}E_{i}g_{1}\left(  E\right)  g_{2}\left(  E\right)  \frac{d}{dE_{i}%
}\sqrt{\left(  \frac{E_{i}^{2}}{g_{2}^{2}\left(  E\right)  }-m_{i}^{2}\left(
r\right)  \right)  ^{3}}dE_{i},\label{Lambda}%
\end{equation}
where%
\begin{equation}
\left\{
\begin{array}
[c]{c}%
m_{1}^{2}\left(  r\right)  =\dfrac{6}{r^{2}}\left(  1-\dfrac{b\left(
r\right)  }{r}\right)  +\dfrac{3}{2r^{2}}b^{\prime}\left(  r\right)
-\dfrac{3}{2r^{3}}b\left(  r\right)  \\
m_{2}^{2}\left(  r\right)  =\dfrac{6}{r^{2}}\left(  1-\dfrac{b\left(
r\right)  }{r}\right)  +\dfrac{1}{2r^{2}}b^{\prime}\left(  r\right)
+\dfrac{3}{2r^{3}}b\left(  r\right)
\end{array}
\right.  \label{masses}%
\end{equation}
and $E^{\ast}$ is the value that leads to a vanishing argument of the square
root. To be more explicit and to evaluate Eqs.$\left(  \ref{rho}%
,\ref{Lambda}\right)  $, we need to specify the form of $g_{1}\left(
E/E_{P}\right)  $ and $g_{2}\left(  E/E_{P}\right)  .$ Following the ansatz of
\cite{GaMa1} we choose:%
\begin{equation}
g_{1}\left(  \frac{E}{E_{P}}\right)  =(1+\beta\frac{E}{E_{P}}+\delta
\frac{E^{2}}{E_{P}^{2}}+\gamma\frac{E^{3}}{E_{P}^{3}})\exp(-\alpha\frac{E^{2}%
}{E_{P}^{2}})\qquad g_{2}\left(  E/E_{P}\right)  =1.\label{g1}%
\end{equation}
The aim of this paper is to study the effective metric induced by quantum
fluctuations of the gravitational field obtained in Eq.$\left(  \ref{diffbr}%
\right)  $. We argue that this effective rainbow metric presents some relevant
novelties with respect to the original (tree-level) rainbow metric mainly due
to the fact that the Planck scale corrections become significant not only for
high energy particles but can become significant also for low-energy
particles. We can recognize two relevant cases:

\begin{description}
\item[a)] $m_{1}^{2}\left(  r\right)  =-m_{2}^{2}\left(  r\right)  =m_{0}%
^{2}\left(  r\right)  ,$

\item[b)] $m_{1}^{2}\left(  r\right)  =m_{2}^{2}\left(  r\right)  =m_{0}%
^{2}\left(  r\right)  .$
\end{description}

When condition a) is satisfied (for example in the Schwarzschild,
Schwarzschild-de Sitter and Schwarzschild -anti-de Sitter cases in proximity
of the throat), Eq.$\left(  \ref{Lambda}\right)  $ can be rearranged in the
following way:%
\begin{equation}%
\begin{array}
[c]{c}%
\dfrac{\Lambda}{8\pi GE_{P}^{4}}=\dfrac{1}{16\pi^{2}\alpha^{7/2}}%
e^{-x^{2}\alpha}\left\{  -\sqrt{\pi}\left[  15\gamma+4x^{2}\alpha^{2}\left(
\beta+x^{2}\gamma\right)  +6\alpha\left(  \beta+2x^{2}\gamma\right)  \right]
-2\alpha^{5/2}\delta e^{\frac{x^{2}\alpha}{2}}\left(  1+e^{x^{2}\alpha
}\right)  x^{4}K_{0}\left(  \dfrac{x^{2}\alpha}{2}\right)  \right. \\
+e^{2x^{2}\alpha}\sqrt{\pi}\left[  2\alpha\left(  -3+2x^{2}\alpha\right)
\beta+\left[  -15-4x^{2}\alpha\left(  -3+x^{2}\alpha\right)  \right]
\gamma\right]  erfc\left(  x\sqrt{\alpha}\right)  +\\
\left.  +2e^{x^{2}\alpha}x\sqrt{\alpha}\left(  -6\alpha\beta-15\gamma
+2x^{2}\alpha\gamma+2\alpha xK_{1}\left(  \dfrac{x^{2}\alpha}{2}\right)
\left[  -2(\alpha+2\delta)\cosh\left(  \dfrac{x^{2}\alpha}{2}\right)
+x^{2}\alpha\delta\sinh\left(  \dfrac{x^{2}\alpha}{2}\right)  \right]
\right)  \right\}
\end{array}
\label{LambdaOM}%
\end{equation}
where $x=\sqrt{m_{0}^{2}\left(  r\right)  /E_{P}^{2}}$ , $\beta_{1}\equiv
\beta$ and where $K_{0}\left(  x\right)  $ and $K_{1}\left(  x\right)  $ are
the Bessel functions and $erf\left(  x\right)  $ is the error function. On the
other hand, when condition b) is satisfied (for example in Minkowski, de
Sitter and anti-de Sitter cases), Eq.$\left(  \ref{Lambda}\right)  $ becomes%
\begin{equation}%
\begin{array}
[c]{c}%
\dfrac{\Lambda}{8\pi GE_{P}^{4}}=\dfrac{1}{8\pi^{2}\alpha^{7/2}}%
e^{-x^{2}\alpha}\left\{  -\sqrt{\pi}\left[  15\gamma+4x^{2}\alpha^{2}\left(
\beta+x^{2}\gamma\right)  +6\alpha\left(  \beta+2x^{2}\gamma\right)  \right]
+\right. \\
\left.  +2e^{\frac{x^{2}\alpha}{2}}x^{2}\alpha^{3/2}\left[  -\alpha\delta
x^{2}K_{0}\left(  \dfrac{x^{2}\alpha}{2}\right)  -\left(  4\delta
+\alpha\left(  2+x^{2}\delta\right)  \right)  K_{1}\left(  \dfrac{x^{2}\alpha
}{2}\right)  \right]  \right\}  .
\end{array}
\label{LambdaEM}%
\end{equation}

Even though we have all orders equations $\left(  \ref{LambdaOM}\right)  $ and
$\left(  \ref{LambdaEM}\right)  $, more insight can be gained considering the
limiting cases. For the purposes of our analysis the case of interests are
those of $x\ll1$ and $x\gg1$. For the case $x\gg1$ the expression $\left(
\ref{LambdaOM}\right)  $\textbf{\ }reduces to\textbf{:}%
\begin{equation}
\frac{\Lambda}{8\pi GE_{P}^{4}}=A_{1}x+{A}_{2}x^{-1}{+}O\left(  x^{-3}\right)
\label{LambdaOMxg}%
\end{equation}
where%
\begin{equation}
A_{1}=-\frac{\left(  2\sqrt{\pi}\alpha^{3/2}+4\alpha\beta+8\gamma+3\sqrt{\pi
}\sqrt{\alpha}\delta\right)  }{8\pi^{2}\alpha^{3}}%
\end{equation}
and%
\begin{equation}
A_{2}=-\frac{16\alpha\beta+48\gamma+3\sqrt{\pi}\sqrt{\alpha}(2\alpha+5\delta
)}{32\pi^{2}\alpha^{4}}.
\end{equation}
It is straightforward to see that the coefficient $A_{1}$ must be set to zero
to have a finite result. Instead for $x\ll1$ we get:%
\begin{equation}
\frac{\Lambda}{8\pi GE_{P}^{4}}=B_{1}+\left[  B_{2}-\frac{1}{8\pi^{2}}%
\ln\left(  \frac{\alpha x^{2}}{4}\right)  \right]  x^{4}+O(x^{5})
\label{LambdaOMxpi}%
\end{equation}
where%
\begin{equation}
B_{1}=-\frac{8\alpha^{3/2}+6\sqrt{\pi}\alpha\beta+15\sqrt{\pi}\gamma
+16\sqrt{\alpha}\delta}{8\pi^{2}\alpha^{7/2}}%
\end{equation}
and%
\begin{equation}
B_{2}=-\frac{(1+2\gamma_{E})\alpha^{3/2}-2\sqrt{\pi}\alpha\beta-\sqrt{\pi
}\gamma-2\sqrt{\alpha}\delta}{16\pi^{2}\alpha^{3/2}}.
\end{equation}
Concerning Eq.$\left(  \ref{LambdaEM}\right)  $ for $x\gg1$ we get:%
\begin{equation}
\frac{\Lambda}{8\pi GE_{P}^{4}}=-e^{-x^{2}\alpha}\left(  \frac{\gamma}%
{2\pi^{3/2}\alpha^{3/2}}x^{4}+\frac{\delta}{2\pi^{3/2}\alpha^{3/2}}%
x^{3}\right)  +O(e^{-x^{2}\alpha}x^{2}), \label{LambdaEMxg}%
\end{equation}

whereas for $x\ll1$ we find:%
\begin{equation}
\frac{\Lambda}{8\pi GE_{P}^{4}}=C_{1}+C_{2}x^{2}+\left[  C_{3}-\frac{1}%
{8\pi^{2}}\ln\left(  x^{2}\alpha/4\right)  \right]  x^{4}+O(x^{5})
\label{LambdaEMxpi}%
\end{equation}

where:%
\begin{equation}
C_{1}=\frac{-8\alpha^{3/2}-6\sqrt{\pi}\alpha\beta-15\sqrt{\pi}\gamma
-16\sqrt{\alpha}\delta}{8\pi^{2}\alpha^{7/2}},
\end{equation}%
\begin{equation}
C_{2}=+\frac{4\alpha^{3/2}+2\sqrt{\pi}\alpha\beta+3\sqrt{\pi}\gamma
+4\sqrt{\alpha}\delta}{8\pi^{2}\alpha^{5/2}}%
\end{equation}
and%
\begin{equation}
C_{3}=\frac{-\alpha^{3/2}-2\gamma_{E}\alpha^{3/2}+2\sqrt{\pi}\alpha\beta
+\sqrt{\pi}\gamma+2\sqrt{\alpha}\delta}{16\pi^{2}\alpha^{3/2}}.
\end{equation}
We can fix our attention on particular forms of spherically-symmetric metrics:
the Schwarzschild geometry and the de Sitter geometry. We begin with the
Schwarzschild geometry.

\section{The Schwarzschild case}

\label{S3}The Schwarzschild's metric is described by $b(r)=R_{S}=2GM$ and from
Eq.$\left(  \ref{masses}\right)  $ we obtain%
\begin{equation}
\left\{
\begin{array}
[c]{c}%
m_{1}^{2}(r)=\frac{6}{r^{2}}\left(  1-\frac{R_{S}}{r}\right)  -\frac{3R_{S}%
}{2r^{3}}=\frac{6}{r^{2}}\left(  1-\frac{5}{4}\frac{R_{S}}{r}\right) \\
\\
m_{2}^{2}(r)=\frac{6}{r^{2}}\left(  1-\frac{R_{S}}{r}\right)  +\frac{3R_{S}%
}{2r^{3}}=\frac{6}{r^{2}}\left(  1-\frac{3}{4}\frac{R_{S}}{r}\right)
\end{array}
\right.  . \label{smasses}%
\end{equation}
If we restrict our attention to the range $R_{S}<r<5R_{S}/4$, which will be
denoted as a \textquotedblleft\textit{short range}\textquotedblright%
\ approximation (SR), we fall into the case a) and we find
\begin{equation}
m_{1}^{2}(r)=-m_{2}^{2}(r)=-m_{0}^{2}(r)=-\frac{3R_{S}}{2r^{3}}%
\end{equation}
with%
\begin{equation}
x=\sqrt{\frac{m_{0}^{2}(r)}{E_{P}^{2}}}=\left(  \frac{3}{2}\frac{R_{S}%
L_{P}^{2}}{r^{3}}\right)  ^{1/2}%
\end{equation}
where we have introduced the Planck length $L_{P}$. For small $x$, Eq.$\left(
\ref{LambdaOMxpi}\right)  $ becomes%
\begin{equation}
\frac{\Lambda}{8\pi GE_{P}^{4}}=B_{1}+\left[  B_{2}-\frac{1}{8\pi^{2}}%
\ln\left(  \frac{3\alpha}{8}\frac{R_{S}L_{P}^{2}}{r^{3}}\right)  \right]
\left(  \frac{3}{2}\frac{R_{S}L_{P}^{2}}{r^{3}}\right)  ^{2}+O\left(
\frac{R_{S}L_{P}^{2}}{r^{3}}\right)  ^{5/2}. \label{ok1}%
\end{equation}
In particular, using the values $\alpha=1/4$, $\beta=-2/(3\sqrt{\pi})$,
$\gamma=\delta=0$ found in\cite{GaMa1}, the parameters $B_{1}$ and $B_{2}$ in
Eq.$\left(  \ref{LambdaOMxpi}\right)  $ become $B_{1}=0$ and $B_{2}=-\left(
9+2\gamma_{E}\right)  /\left(  16\pi^{2}\right)  $ so that the effective
metric $\left(  \ref{effbr}\right)  $ reduces to%
\begin{equation}
b_{eff}(r)=R_{S}+\frac{3R_{S}^{2}L_{P}^{2}}{4\pi r^{3}}\left[  \ln\left(
\frac{3}{32}\frac{R_{S}L_{P}^{2}}{r^{3}}\right)  +\gamma_{E}+\frac{7}%
{2}\right]  -\frac{3L_{P}^{2}}{4\pi R_{S}}\left[  \ln\left(  \frac{3}{32}%
\frac{L_{P}^{2}}{R_{S}^{2}}\right)  +\gamma_{E}+\frac{7}{2}\right]  +O\left(
\frac{L_{P}^{3}}{R_{S}^{2}}\right)  . \label{ric1}%
\end{equation}
Note that the use of expression $\left(  \ref{LambdaOMxpi}\right)  $ is
appropriate whenever $R_{S}\gg L_{P}$. However in case $B_{1}$ was not nought,
then%
\[
b_{eff}(r)\simeq R_{S}+8\pi B_{1}\frac{r^{3}}{L_{P}^{2}}+\frac{3R_{S}^{2}%
L_{P}^{2}}{4\pi r^{3}}\left[  \ln\left(  \frac{3}{32}\frac{R_{S}L_{P}^{2}%
}{r^{3}}\right)  +\gamma_{E}+\frac{7}{2}\right]
\]%
\begin{equation}
-\frac{3L_{P}^{2}}{4\pi R_{S}}\left[  \ln\left(  \frac{3}{32}\frac{L_{P}^{2}%
}{R_{S}^{2}}\right)  +\gamma_{E}+\frac{7}{2}\right]  -8\pi B_{1}\frac
{R_{S}^{3}}{L_{P}^{2}}+O\left(  \frac{L_{P}^{3}}{R_{S}^{2}}\right)  .
\end{equation}
Another interesting case is $x\gg1$ or $L_{P}\gg R_{S}$ (sub-Planckian
wormhole). The correct expression to use is given by Eq.$\left(
\ref{LambdaOMxg}\right)  $ and the effective metric becomes%
\begin{equation}
b_{eff}(r)=R_{S}+\frac{32}{9\sqrt{\pi}L_{P}^{2}}\left(  \frac{2}{3R_{S}%
L_{P}^{2}}\right)  ^{1/2}r^{9/2}+O\left(  r^{9/2}R_{S}^{-1/2}L_{P}%
^{-3}\right)  .
\end{equation}
We now fix our attention on the other range of approximation, namely when
$r\gg R_{S}$. In this case we are in the \textquotedblleft\textit{long
range}\textquotedblright\ approximation (LR) and the effective masses $\left(
\ref{smasses}\right)  $ are distinct. In this approximation both $m_{1}(r)$
and $m_{2}(r)$ are positive that it means that we are dealing with a
$\Lambda(x)$ given by Eq.$\left(  \ref{LambdaEM}\right)  $%
\begin{equation}
\frac{\Lambda(x_{1},x_{2})}{8\pi GE_{P}^{4}}=\frac{\Lambda(x_{1})}{2}%
+\frac{\Lambda(x_{2})}{2},
\end{equation}
where%
\begin{equation}
x_{1}=\sqrt{\frac{m_{1}^{2}(r)}{E_{P}^{2}}}\text{\qquad\textrm{and\qquad}%
}x_{2}=\sqrt{\frac{m_{2}^{2}(r)}{E_{P}^{2}}}.
\end{equation}
Since we are in the LR approximation, we can claim that $x_{1},x_{2\text{ }%
}\ll1$ and we can use the expansion of Eq.$\left(  \ref{LambdaEMxpi}\right)
$. Thus explicitly:%
\begin{equation}
\frac{\Lambda}{8\pi GE_{P}^{4}}=C_{1}+\frac{C_{2}}{2}\left(  x_{1}^{2}%
+x_{2}^{2}\right)  +\frac{C_{3}}{2}\left(  x_{1}^{4}+x_{2}^{4}\right)
-\frac{1}{16\pi^{2}}\left[  \ln\left(  x_{1}^{2}\alpha/4\right)  x_{1}^{4}%
+\ln\left(  x_{2}^{2}\alpha/4\right)  x_{2}^{4}\right]  +O(x_{1}^{5},x_{2}%
^{5})
\end{equation}

Setting properly the parameters i.e. using the values $\alpha=1/4$,
$\beta=-2/(3\sqrt{\pi})$ and adopting the simple choice $\gamma=\delta=0$ we
would obtain at the leading order $\Lambda(r)\simeq1/r^{2}$ and thus
$b_{eff}(r)\simeq R_{S}+64/\pi r$ i.e. we would get an increasing large
$b_{eff}(r)$ that would be incompatible with observations. To maintain
compatibility with observations we have to request that also $C_{2}=0$ that
fixes the parameters $\gamma=2/(9\sqrt{\pi})$ and $\delta=-5/12$. This choice
leads to%
\begin{equation}
\frac{\Lambda(r)}{8\pi GE_{P}^{4}}=\frac{36}{\pi^{2}}\ln\left(  \dfrac
{r}{L_{P}}\right)  \left(  \frac{L_{P}^{2}}{r^{2}}\right)  ^{2}+O\left(
\frac{L_{P}^{2}}{r^{2}}\right)  ^{2}%
\end{equation}
which, in terms of $b_{eff}(r)$ means%
\begin{equation}
b_{eff}(r)=R_{S}+\frac{288}{\pi}\frac{L_{P}^{2}}{r}\ln\left(  \frac{r}{L_{P}%
}\right)  +O\left(  \frac{L_{P}^{2}}{r}\right)  .\label{b(r)larger}%
\end{equation}
The different expressions of the effective metric in the different r\'{e}gimes
can be used to see the effects on a test particle moving on a background of
the form $\left(  \ref{metric}\right)  $. Even in this case, we will fix our
attention on the Schwarzschild and the de Sitter space-time, respectively. We
begin with the Schwarzschild background.

\section{Implications for test particle on a Schwarzschild background}

\label{S4} Using the appropriate form of the effective metric, we consider the
modifications induced on the orbital motion and onto the effective potential.

\subsection{The orbital motion}

In Gravity's Rainbow the motion is still geodetic but the geodetic equation
(see \cite{Ling:2006ba,Leiva:2008fd}) acquires, by means of the connection,
the dependency on the energy $E$ of the test particle:%
\begin{equation}
\frac{d^{2}x^{\mu}}{d\tau^{2}}+\Gamma_{\alpha,\beta}^{\mu}(E)\frac{dx^{\alpha
}}{d\tau}\frac{dx^{\beta}}{d\tau}=0.
\end{equation}

It is easy to check that the quantity%
\begin{equation}
g_{\mu\nu}(E)\frac{dx^{\alpha}}{d\tau}\frac{dx^{\beta}}{d\tau}=-\mu
\label{motionconst}%
\end{equation}
defines a constant of the motion, being $\mu=1$ for massive particles
(time-like geodesics) and $\mu=0$ for massless particles (null-like
geodesics). Using $\left(  \ref{metric}\right)  $ with $\left(
\ref{motionconst}\right)  $ provide us the following equation%
\begin{equation}
-\exp[-2\Phi(r)]\frac{\dot{t}^{2}}{g_{1}^{2}\left(  E\right)  }+\frac{\dot
{r}^{2}}{\left(  1-\frac{b_{eff}\left(  r\right)  }{r}\right)  g_{2}%
^{2}\left(  E\right)  }+\frac{r^{2}}{g_{2}^{2}\left(  E\right)  }\left[
\dot{\theta}^{2}+\sin^{2}\theta\dot{\phi}^{2}\right]  =-\mu
.\label{motionconst2}%
\end{equation}

Assuming an equatorial plane $\theta=\pi/2$ and using the following constants
of the motion%
\begin{align}
\epsilon &  =\frac{\exp[-2\Phi(r)]}{g_{1}^{2}\left(  E\right)  }\dot{t}\\
l &  =r^{2}\dot{\phi},\nonumber
\end{align}
Eq.$\left(  \ref{motionconst2}\right)  $ can be written in the form:%
\begin{equation}
\frac{\epsilon^{2}g_{1}^{2}\left(  E\right)  g_{2}^{2}\left(  E\right)
}{\left(  1-\frac{b_{eff}\left(  r\right)  }{r}\right)  }-\frac{\dot{r}^{2}%
}{1-\frac{b_{eff}\left(  r\right)  }{r}}-\frac{l^{2}}{r^{2}}=g_{2}^{2}\left(
E\right)  \mu,\label{intP}%
\end{equation}
where we have used the relation\textbf{ }$\exp[-\Phi(r)]=\sqrt{1-b\left(
r\right)  /r}$ for the Schwarzschild case.\textbf{ }Then in term of the
variable $u=1/r$, being $\dot{r}=-ldu/d\phi$ and differentiating with respect
to $\phi$, we obtain the equation of the orbit%
\begin{equation}
\frac{d^{2}u}{d\phi^{2}}+u-\frac{b_{eff}\left(  u^{-1}\right)  }{2}\left[
3u^{2}+\frac{\mu g_{2}^{2}\left(  E\right)  }{l^{2}}\right]  +\frac
{b_{eff}^{\prime}(u^{-1})}{2}\left[  u+\frac{\mu g_{2}^{2}\left(  E\right)
}{ul^{2}}\right]  =0.\label{orbmot}%
\end{equation}

We can evaluate Eq.$\left(  \ref{orbmot}\right)  $ using the form of $b_{eff}$
given by $\left(  \ref{b(r)larger}\right)  $. The leading corrections to the
usual general relativistic result are of order of $\delta=L_{P}^{2}u/R_{S}%
\ln\left(  uL_{P}\right)  $. For the planets of the solar system one gets the
greatest correction for Mercury orbit\textbf{:} $\delta\simeq10^{-82}$ that it
means that the quantum effects are largely negligible. Notice that in the case
of photons, being $\mu=0$\textbf{,} the Binet equation takes the form%
\begin{equation}
\frac{d^{2}u}{d\phi^{2}}+u\left[  1-\frac{3}{2}b_{eff}\left(  u^{-1}\right)
u\right]  +\frac{u}{2}b_{eff}^{\prime}=0,
\end{equation}
so that quantum corrections modify the orbit only through the loop effects
(i.e. by the modification induced in $b_{eff}$). At the tree-level, being
$b_{eff}=R_{S}$ and $b_{eff}^{\prime}=0,$ there are no corrections to the
orbit with respect to the standard general relativistic case, in agreement
with what is found in Ref.\cite{Leiva:2008fd}.

\subsection{Photon time delay in radial motion.}

For massless particles $\left(  \mu=0\right)  $ in radial motion using
equation $\left(  \ref{motionconst2}\right)  $ one obtains%
\begin{equation}
\frac{dr}{dt}=\exp[-\Phi(r)]\frac{g_{2}\left(  E\right)  }{g_{1}\left(
E\right)  }\sqrt{1-\frac{b_{eff}\left(  r\right)  }{r}}.
\end{equation}
Now let us consider two photons, both emitted at $r=r_{e}$ and detected at
$r=r_{d}$. The first photon be an high-energy photon, the second one be a
low-energy photon $\left(  E\approx0\right)  $. Then the time delay of
detection between the two is given by%
\begin{equation}
\Delta t=\left[  \frac{g_{1}\left(  E\right)  }{g_{2}\left(  E\right)
}-1\right]
{\displaystyle\int\limits_{r_{e}}^{r_{d}}}
\frac{\exp[\Phi(r)]}{\sqrt{1-\dfrac{b_{eff}\left(  r\right)  }{r}}%
}dr.\label{dtsfero}%
\end{equation}

This last formula holds for a general spherical symmetric metric of the type
$\left(  \ref{metric}\right)  $. In particular for the Schwarzschild case,
being $b_{eff}\left(  r\right)  =R_{S}+b_{qc}\left(  r\right)  $, Eq.$\left(
\ref{dtsfero}\right)  $ takes the form%
\begin{equation}
\Delta t\simeq\Delta t_{1}+\Delta t_{2},\label{dtS}%
\end{equation}
where%
\begin{align}
\Delta t_{1} &  =\left[  \frac{g_{1}\left(  E\right)  }{g_{2}\left(  E\right)
}-1\right]
{\displaystyle\int\limits_{r_{e}}^{r_{d}}}
\frac{dr}{1-\dfrac{R_{S}}{r}},\label{dtS1}\\
\Delta t_{2} &  =\left[  \frac{g_{1}\left(  E\right)  }{g_{2}\left(  E\right)
}-1\right]
{\displaystyle\int\limits_{r_{e}}^{r_{d}}}
\frac{b_{qc}\left(  r\right)  dr}{r\left(  1-\dfrac{R_{S}}{r}\right)  ^{2}%
},\label{dtS2}%
\end{align}
and where we have neglected terms of order $b_{qc}^{2}\left(  r\right)  $ or
higher. Formula $\left(  \ref{dtS1}\right)  $ can be easily evaluated. Being
in our assumptions, at the lowest order in the Planck scale, $g_{2}\left(
E\right)  =1,$ $g_{1}\left(  E\right)  \simeq1+\beta E/E_{P}$ and
$\beta=-2/(3\sqrt{\pi}),$ we get%
\begin{equation}
\Delta t_{1}=-\frac{2}{3\sqrt{\pi}}\frac{E}{E_{P}}r_{d}\left[  1-\frac{r_{e}%
}{r_{d}}+\frac{R_{S}}{r_{d}}\ln\left(  \frac{r_{d}/R_{S}-1}{r_{e}/R_{S}%
-1}\right)  \right]  .
\end{equation}
To analytically evaluate the integral $\left(  \ref{dtS2}\right)  $ we have to
specify the exact form of $b_{QC}\left(  r\right)  .$ For instance, under the
hypothesis of validity of Eq.$\left(  \ref{b(r)larger}\right)  $, we get%
\begin{equation}
\Delta t_{2}\simeq-\frac{2}{3\sqrt{\pi}}\frac{288}{\pi^{2}}\frac{E}{E_{P}%
}L_{P}^{2}\left[  \frac{1}{r_{e}}-\frac{1}{r_{d}}+\frac{1}{r_{e}}\ln\left(
\frac{r_{e}}{L_{P}}\right)  -\frac{1}{r_{d}}\ln\left(  \frac{r_{d}}{L_{P}%
}\right)  \right]  .
\end{equation}

Notice that, whereas the tree level contribution grows with the distance
between the detector and the source, the quantum fluctuations given by $\Delta
t_{2}$ are more and more suppressed as the distance from the source increases.
To provide a numerical estimation, if we considered two photons emitted at the
same time in $r_{e}\simeq10km$ from a star in the center of our galaxy
$\left(  r_{d}\simeq3\cdot10^{4}ly\right)  ,$ we would have $\Delta
t(\simeq\Delta t_{1})\simeq1s$ already with $E_{2}\simeq10^{-12}E_{P}$, and
the effects of quantum fluctuations, being $\Delta t_{2}\simeq10^{-86}s$,
would be largely negligible.

\subsection{The effective potential}

We can also study the modifications induced on the effective potential by the
quantum fluctuation of the metric with respect to the classical general
relativistic case. Let us notice that by the standard procedure Eq.$\left(
\ref{intP}\right)  $ can be rearranged in the form%
\begin{equation}
\dot{r}^{2}+V^{2}(r)=\epsilon^{2}g_{1}^{2}\left(  E\right)  g_{2}^{2}\left(
E\right)  ,
\end{equation}
where now%
\begin{equation}
V(r)=\sqrt{\left(  g_{2}^{2}\left(  E\right)  +\frac{l^{2}}{r^{2}}\right)
\left(  1-\frac{b_{eff}\left(  r\right)  }{r}\right)  }=V_{GR}(r)\sqrt
{1-\frac{b_{qc}\left(  r\right)  }{r-R_{S}}}, \label{veff}%
\end{equation}
and where
\begin{equation}
V_{GR}(r)=\sqrt{\left(  1+\frac{l^{2}}{r^{2}}\right)  \left(  1-\frac{R_{S}%
}{r}\right)  }%
\end{equation}
is the classical general-relativistic effective potential. In Eq.$\left(
\ref{veff}\right)  $ we have assumed $g_{2}\left(  E\right)  =1$ and we have
defined $b_{eff}\left(  r\right)  =R_{S}+b_{qc}\left(  r\right)  $. To
explicitly evaluate $\left(  \ref{veff}\right)  $ we have to introduce
$b_{eff}\left(  r\right)  $ in the formula. For instance, in the range in
which Eq.$\left(  \ref{b(r)larger}\right)  $ holds $(r\gg R_{S})$ we find%
\begin{equation}
V(r)\simeq V_{GR}(r)\left[  1-\frac{144}{\pi}\frac{L_{P}^{2}\ln\left(
r/L_{P}\right)  }{r^{2}}\right]  .
\end{equation}

We notice that Planck-scale corrections begin to manifest at the one-loop
level, as far as space-time is probed at Planckian scales (i.e. at scales at
which the terms in $L_{P}/r$ become relevant). It is also interesting to note
here that if one considered the scattering between two masses in a
gravitational potential, at the lowest order in Planck scale, one would find,
according to Ref.\cite{Donoghue:1994dn}, the Donoghue's potential%
\begin{equation}
V_{Donoghue}(r)=-\frac{GM}{r}\left[  1-\frac{G(M+m)}{rc^{2}}-\frac
{127L_{P}^{2}}{30\pi^{2}r^{2}}\right]  .
\end{equation}

In the Donoghue's potential the leading quantum corrections are of the form
$L_{P}^{2}/r^{2}$, while in our case they are of order $L_{P}^{2}/r^{2}%
\ln\left(  r/L_{P}\right)  $. Once the analysis has been made for the
Schwarzschild background it is not difficult to extend it to other spherically
symmetric cases. In the next section, we will analyze the de Sitter background.

\section{The de Sitter case}

\label{S5} The de Sitter case written in static coordinates is simply
described by $b\left(  r\right)  =\Lambda r^{3}/3$. In this situation the
effective masses of Eq.$\left(  \ref{masses}\right)  $ take the form%
\begin{equation}
m_{1}^{2}\left(  r\right)  =m_{2}^{2}\left(  r\right)  =\frac{6}{r^{2}%
}-\Lambda,\qquad r\in\left(  0,r_{C}\right]
\end{equation}
with $r_{C}=\sqrt{3/\Lambda}$. Defining the dimensionless variable%
\begin{equation}
x=\frac{L_{P}}{r}\sqrt{6-\Lambda r^{2}},
\end{equation}
we can use expansion $\left(  \ref{LambdaEMxpi}\right)  $, assuming $r\gg
L_{P}$ and $\Lambda r^{2}=O(1),$ to obtain%
\begin{equation}
\frac{\Lambda}{8\pi GE_{P}^{4}}=C_{1}+C_{2}\left(  \frac{L_{P}}{r}\right)
^{2}\left(  6-\Lambda r^{2}\right)  +\left\{  C_{3}-\frac{1}{8\pi^{2}}%
\log\left[  \left(  \frac{L_{P}}{r}\right)  ^{2}\left(  6-\Lambda
r^{2}\right)  ^{2}\alpha/4\right]  \right\}  \left(  \frac{L_{P}}{r}\right)
^{4}\left(  6-\Lambda r^{2}\right)  ^{2}+O\left(  \frac{L_{P}}{r}\right)
^{5}.
\end{equation}

Again substituting in Eq.$\left(  \ref{diffbr}\right)  $ we get the
quantum-corrected de Sitter parameter%
\begin{align}
b_{eff}(r) &  =\frac{\Lambda+C_{1}}{3}r^{3}+r\left(  6-\frac{1}{3}r^{2}%
\Lambda\right)  C_{2}L_{P}^{2}+\label{beffDS}\\
&  +\frac{L_{P}^{4}}{r}\left\{  \frac{9}{\pi^{2}}\log\left(  \frac{3L_{P}%
\sqrt{\alpha}}{r}\right)  -9\left(  4C_{3}+\frac{1}{\pi^{2}}\right)
+r^{2}\Lambda\left[  -12C_{3}+\frac{9}{2\pi^{2}}+\frac{3}{\pi^{2}}\log\left(
\frac{3L_{P}\sqrt{\alpha}}{r}\right)  \right]  \right\}  +O[\Lambda
]^{2}.\nonumber
\end{align}

If we assume compatibility with the Schwarzschild-background case then we have
to set $C_{1}\simeq C_{2}\simeq0$ so that at the leading order we would have%
\begin{equation}
\Lambda_{eff}\simeq\Lambda+27\frac{L_{P}^{4}}{r^{4}}\left\{  \frac{1}{\pi^{2}%
}\log\left(  \frac{3L_{P}\sqrt{\alpha}}{r}\right)  -\left(  4C_{3}+\frac
{1}{\pi^{2}}\right)  +r^{2}\Lambda\left[  -12C_{3}+\frac{9}{2\pi^{2}}+\frac
{3}{\pi^{2}}\log\left(  \frac{3L_{P}\sqrt{\alpha}}{r}\right)  \right]
\right\}  +O[\Lambda^{2},L_{P}^{5}/r^{5}]. \label{Lambda_eff_dS1}%
\end{equation}

If instead we admit the possibility $C_{1}$ and $C_{2}$ to assume values that
are different from those assumed on the Schwarzschild background, at the
leading order we would find%
\begin{equation}
\Lambda_{eff}\simeq\Lambda+C_{1}+C_{2}\frac{L_{P}^{2}}{r^{2}}\left(
18-r^{2}\Lambda\right)  . \label{Lambda_eff_dS2}%
\end{equation}

Notice that from Eqs.$\left(  \ref{Lambda_eff_dS1}-\ref{Lambda_eff_dS2}%
\right)  $ follows that even starting from exactly $\Lambda=0,$ one can obtain
a nonvanishing cosmological constant induced by the quantum fluctuations,
depending on the parameters of the rainbow functions. Thus we could have an
effective de Sitter space-time starting from a bare Minkowski space-time. We
can also have the situation where quantum fluctuations cancel the nonvanishing
"\textit{bare cosmological constant}" $\left(  \Lambda\neq0\right)  $
providing $\Lambda_{eff}\simeq0$. In this last case we would have an effective
Minkowski space-time starting from a bare de Sitter space-time. Finally it
could be also possible that quantum fluctuations be able to transform a de
Sitter space-time into an anti-de Sitter space-time or viceversa. All these
cases seem to suggest how ZPE be a source of a topology change\cite{DGL}%
.\textbf{\ }

\subsection{Implications for photons propagating on a de Sitter background}

To study the motion of photons on a de Sitter background one can follow the
strategy outlined in the previous section for the Schwarzschild metric. In
particular photon time delay for radial motion can be by inferred directly
from Eq.$\left(  \ref{dtsfero}\right)  $ with the simple prescription of using
the form of $b_{eff}\left(  r\right)  $ given by Eq.$\left(  \ref{beffDS}%
\right)  $. However we observe here that the metric%
\begin{equation}
ds^{2}=-\left(  1-\frac{\Lambda_{eff}}{3}r^{2}\right)  \frac{dt^{2}}{g_{1}%
^{2}\left(  E\right)  }+\frac{dr^{2}}{\left(  1-\frac{\Lambda_{eff}}{3}%
r^{2}\right)  g_{2}^{2}\left(  E\right)  }+\frac{r^{2}}{g_{2}^{2}\left(
E\right)  }\left(  d\theta^{2}+\sin^{2}\theta d\phi^{2}\right)  , \label{dsm}%
\end{equation}
is expressed in terms of "static coordinates" while for our purposes it is
better to write it in terms of "flat coordinates". Indeed flat coordinates
appear more natural from a phenomenological perspective since they allow to
associate the motion of a detector to a given-comoving position. The change
between "static" and "flat" coordinates can be obtained in Gravity's Rainbow
by means of the map%
\begin{align}
t^{\prime}  &  =t+\frac{1}{2\sqrt{\Lambda_{eff}/3}}\frac{g_{1}\left(
E\right)  }{g_{2}\left(  E\right)  }\log\left(  1-\Lambda_{eff}r^{2}/3\right)
,\label{map1}\\
\rho &  =\frac{r}{g_{2}\left(  E\right)  }\exp\left(  -t^{\prime}\sqrt
{\Lambda_{eff}/3}\frac{g_{2}\left(  E\right)  }{g_{1}\left(  E\right)
}\right)  . \label{map2}%
\end{align}

In terms of the variables $\rho=\sqrt{x^{2}+y^{2}+z^{2}},t^{\prime}$ given by
Eqs.$\left(  \ref{map1}-\ref{map2}\right)  $, the metric $\left(
\ref{dsm}\right)  $ becomes%
\begin{equation}
ds^{2}=-\frac{dt^{\prime2}}{g_{1}^{2}\left(  E\right)  }+\exp\left[
2\frac{g_{2}\left(  E\right)  }{g_{1}\left(  E\right)  }\sqrt{\Lambda_{eff}%
/3}t\right]  \left(  dx^{2}+dy^{2}+dz^{2}\right)  ,
\end{equation}
from which one can easily deduce the photon equation of motion%
\begin{equation}
x_{dS}(t)=\frac{e^{\frac{\sqrt{\Lambda_{eff}/3}}{g_{1}\left(  E\right)  }%
t_{0}}-e^{-\frac{\sqrt{\Lambda_{eff}/3}}{g_{1}\left(  E\right)  }t}}%
{\sqrt{\Lambda_{eff}/3}},
\end{equation}
where according to $\left(  \ref{g1}\right)  $ we have assumed $g_{2}\left(
E\right)  =1.$

Following Ref.\cite{Marciano:2010gq} let us now consider two photons emitted
at the same time $t=-t_{0}$ at $x_{dS}=0.$ The first photon be a low energy
photon ($E\ll E_{P})$ and the second one be a Planckian photon ($E\sim
E_{P}).$ Both photons are assumed to be detected at a later time in
$\overline{x}_{dS}.$ We expect to detect the two photons with a time delay
$\Delta t$ given by the solution of the equation%
\begin{equation}
\overline{x}_{dS}^{E\ll E_{P}}(0)=\overline{x}_{dS}^{E\sim E_{P}}(\Delta t),
\end{equation}
that implies%
\begin{equation}
\Delta t\simeq g_{1}\left(  E\right)  \frac{e^{\sqrt{\Lambda_{eff}/3}t_{0}%
}-e^{\frac{\sqrt{\Lambda_{eff}/3}}{g_{1}(E)}t_{0}}}{\sqrt{\Lambda_{eff}/3}%
}\simeq\beta\frac{E}{E_{P}}t_{0}\left(  1+\sqrt{\Lambda_{eff}/3}t_{0}\right)
, \label{timed}%
\end{equation}
where we have used $\left(  \ref{g1}\right)  $ for the rainbow functions.

At the lowest order the formula $\left(  \ref{timed}\right)  $ agrees with the
corresponding formula of Ref.\cite{Marciano:2010gq} in which the Planck scale
comes into play by means of a parameter $w=f(HL_{P})$ appearing in the
quantum-de Sitter group. The result for the time delay found in
Ref.\cite{Marciano:2010gq} is%
\begin{equation}
\Delta t\simeq p(1-e^{2Ht_{0}})/(2H^{2})w,\label{eqm}%
\end{equation}
where $H=$ $\sqrt{\Lambda/3}.$ A first difference between $\left(
\ref{timed}\right)  $ and $\left(  \ref{eqm}\right)  $ regards the origin of
the Planck parameter for the time delay that in $\left(  \ref{eqm}\right)  $
is connected to the quantum deformation of the classical de Sitter group
\textquotedblleft$w$\textquotedblright\ whereas in $\left(  \ref{timed}%
\right)  $ comes from the rainbow deformation \textquotedblleft$\beta
$\textquotedblright. A second key difference between $\left(  \ref{timed}%
\right)  $ and $\left(  \ref{eqm}\right)  $ is in $\Lambda_{eff}$. Indeed
according to Eqs.$\left(  \ref{Lambda_eff_dS1}-\ref{Lambda_eff_dS2}\right)  $
the effective cosmological constant includes quantum corrections: this further
dependence on $L_{P}$ is not present in $\left(  \ref{eqm}\right)  $.

\subsection{Quantum corrections to Minkowski space-time}

The Minkowski limit can be derived directly from Eq.$\left(  \ref{beffDS}%
\right)  $ by assuming a vanishing bare cosmological constant $\left(
\Lambda=0\right)  $. In this case one obtains for the space-time metric
parameter%
\begin{equation}
b_{eff}(r)=\frac{C_{1}}{3}r^{3}+6rC_{2}L_{P}^{2}+\frac{9L_{P}^{4}}{r}\left\{
\frac{1}{\pi^{2}}\log\left(  \frac{3L_{P}\sqrt{\alpha}}{r}\right)  -\left(
4C_{3}+\frac{1}{\pi^{2}}\right)  \right\}  ,
\end{equation}
and consequently, for the cosmological constant%
\begin{equation}
\Lambda_{eff}(r)\simeq.C_{1}+18C_{2}\frac{L_{P}^{2}}{r^{2}}+\frac{27L_{P}^{4}%
}{r^{4}}\left\{  \frac{1}{\pi^{2}}\log\left(  \frac{3L_{P}\sqrt{\alpha}}%
{r}\right)  -\left(  4C_{3}+\frac{1}{\pi^{2}}\right)  \right\}  ,
\end{equation}
from which it is easily seen that, at the lowest order, the quantum
corrections to a bare Minkowski space-time can in principle transform it into
either a de Sitter or an anti-de Sitter space-time, depending on the signs and
on the values of the constants $C_{1}$ and $C_{2}$ (which in turn depend on
the parameters of the deformed dispersion relations). If we assume
$C_{1}\simeq C_{2}$ $\simeq0,$ according to what we have found for the
Schwarzschild background, we get%
\begin{equation}
\Lambda_{eff}(r)\simeq.\frac{27L_{P}^{4}}{r^{4}}\left\{  \frac{1}{\pi^{2}}%
\log\left(  \frac{3L_{P}\sqrt{\alpha}}{r}\right)  -\left(  4C_{3}+\frac{1}%
{\pi^{2}}\right)  \right\}  . \label{LMink}%
\end{equation}

From Eq.$\left(  \ref{LMink}\right)  $ follows that quantum corrections are
actually more likely to turn Minkowski space-time it into an anti-de Sitter
space-time rather than into a de Sitter one. To evaluate how rainbow effects
and cosmological-constant-induced quantum fluctuations affect time delay for a
particle in motion in a Minkowski space-time, we can use directly formulas
$\left(  \ref{dtS}-\ref{dtS2}\right)  $ with the prescription of fixing
$R_{S}=0$. It can be interesting here to compare the time delay derived from
$\left(  \ref{dtS}-\ref{dtS2}\right)  $ with the experimental time delay whose
measure has very recently been announced by\cite{OPERA:2011zb}. In
\cite{OPERA:2011zb} a negative time delay $\Delta t_{\exp}\simeq-60ns$ has
been claimed  for $E\simeq17Gev$ neutrinos propagating over a distance
$(r_{d}-r_{e})\simeq730km$ with respect to the time of propagation of light
(i.e. of low-energy photons). Substituting in $\left(  \ref{dtS}%
-\ref{dtS2}\right)  $ the values of the parameters corresponding to the
experimental settings of \cite{OPERA:2011zb}, and treating the neutrinos as
massless particles, we find%
\begin{equation}
\Delta t\simeq\Delta t_{1}\simeq\beta\frac{E}{E_{P}}\left(  r_{d}%
-r_{e}\right)  \approx-10^{-12}ns\approx-10^{-14}\Delta t_{\exp},
\end{equation}
being in our case $\beta=-2/(3\sqrt{\pi})$. Thus our formulas with the value
of $\beta$ fixed according to the procedure followed in \cite{GaMa1} predict a
time delay $10^{-14}$ times smaller than the announced experimental one, and
with the opposite sign (i.e. our formulas predict high-energy neutrinos to be
slower than low-energy photons), neither the quantum fluctuations of the
metric encoded in $\Delta t_{2}(\ll\Delta t_{1})$ are able to qualitatively
change this conclusion. It is worth noticing that in order to account for the
proper sign allowing the superluminal neutrino propagation of
\cite{OPERA:2011zb}, and in order to get the right time delay $\Delta
t\simeq\Delta t_{\exp},$we should assume $\beta\approx+10^{14}$. This
huge-positive and unnatural value of $\beta$ however would also be in contrast
with other experimental data such us those relative to the time delay observed
in TeV flares coming from active galaxies \cite{Biller:1998hg}, that should be
various orders of magnitude greater than measured.

\section{Conclusions}

\label{S6}We have found that in Gravity's Rainbow quantum fluctuations can
significantly modify the structure of the tree level space-time metric and,
consequently, the propagation of test particles. The modifications induced on
the metric depend on the parameters that deform the energy-momentum dispersion
relation. The request of recovering the classical limit at large distances
from the source constrains the classes of admissible deformation of the
dispersion relation. A result that is common to all the spherical symmetric
backgrounds that we have analyzed is that the metric is Planck scale deformed
also in the case one should consider low-energy particles as long as Planck
scale distances are involved. In fact the quantum effects modify the metric in
two different ways. The first way is by means of the dependency on the Planck
scale of the rainbow functions. This type of modification is significant only
as soon as the energy of the test particle approaches the Planck energy. The
second way is through the quantum fluctuations of the metric that in our
approach are finite. This other type of modification becomes significant
independently on the energy of the test particle and manifests on Planck
length scale even for low-energy particles. For the Schwarzschild background
we have found that the quantum effects alter the orbit of the test particles
(even those of massless test particles) and not only the time of propagation,
as instead happens at the tree level (see Ref.\cite{Leiva:2008fd}). The
leading quantum correction at large distances $r$ from a source, whose
Schwarzschild radius is $R_{S}$, is of order $L_{P}^{2}/(R_{S}r)\ln\left(
r/L_{P}\right)  $ with respect to the usual general relativistic correction.
This means that a direct observation of a modification of the orbit is largely
out of reach. Instead the leading effect on the time delay between two photons
being amplified by the distance from the source could be more effective with
respect to the possibility of being experimentally revealed. However the
effect of fluctuations of the metric appear to be largely negligible with
respect to the leading tree level effect given by the rainbow functions.
Finally our calculation of the effective potential has shown that the
gravitational potential is modified (again also for low energy test particles)
but only at Planckian length scales. The type of modifications induced on the
potential is, at the leading order, similar to others already appeared in
literature and can be influent in astrophysical situations in which ultra-high
densities are reached \cite{AmelinoCamelia:2009tv,AmelinoCamelia:2010rm}.
Concerning the de Sitter case our analysis has shown that quantum fluctuations
of the de Sitter metric can greatly influence the effective value of the
cosmological constant. In particular quantum fluctuations could be able to
change a de Sitter space time into an anti-de Sitter one or even into a
Minkowski one. It is also likely to get an effective anti-de Sitter, rather
than a de Sitter, space-time starting from a tree level Minkowski space-time.
The analysis of the time delay of photons in the de Sitter case has lead to a
result that is qualitatively similar to another already appeared in literature
in the different framework of the quantum groups.\textbf{ }The analysis of the
time delay of photons in Minkowski space-time, also accounting for the quantum
fluctuations of the metric, has however lead us to a result that compared with
the experimental delay announced in \cite{OPERA:2011zb} is roughly $14$ orders
of magnitude smaller, and with the opposite sign. Thus we cannot account for
the experimental data reported in \cite{OPERA:2011zb} unless we do not assume
for the parameter deforming the energy/momentum dispersion relation the value
$\beta\approx+10^{14}$, that however (see also \cite{AmelinoCamelia:2011dx})
would remain in contrast with other experimental data as e.g. those of
Ref.\cite{Biller:1998hg}. However, if the OPERA measure reveals correct, we do
not have to forget that our analysis is completely based on wave functionals
that obey Bose-Einstein statistics, while neutrinos obey the Fermi-Dirac
statistics that it means that the $\beta$ parameter can assume the correct
sign to obtain superluminal neutrinos.

\end{document}